%% file: main.tex
\documentclass{sig-alternate-10pt}

\usepackage{epsfig}
\usepackage{graphics}
\usepackage{amssymb,amsmath}
\usepackage{multirow}
\usepackage{color}
\usepackage{listings}
\usepackage{xspace}
\usepackage{subcaption}
\usepackage{caption}
\usepackage[section]{placeins}
\usepackage{url}

\usepackage{breakurl}
\usepackage[breaklinks]{hyperref}
\usepackage{times}
\newcommand{\miniproxy}{Miniproxy\xspace}

\begin{document}

\title{On-the-Fly TCP Acceleration with Miniproxy
\titlenote{This work was partly funded by the EU in the context of the 
\mbox{SUPERFLUIDITY} project (5G PPP).}}

\author{
Giuseppe Siracusano$^{\ddagger}$ $^{\dagger}$, Roberto Bifulco$^{\dagger}$, Simon Kuenzer$^{\dagger}$,\\
Stefano Salsano$^{\ddagger}$, Nicola Blefari Melazzi$^{\ddagger}$, Felipe Huici$^{\dagger}$\\ 
\small{$^{\dagger}$NEC Laboratories Europe, Germany}  
\small{$^{\ddagger}$Univ. of Rome Tor Vergata/CNIT, Italy}\\ 
\\
Extended version of paper published in ACM HotMiddlebox 2016 - May 2016\\
\\[-9.0ex]
}

\maketitle

\begin{abstract}
TCP proxies are basic building blocks for many advanced middleboxes.
In this paper we present Miniproxy, a TCP proxy built on top of a specialized minimalistic cloud operating system. Miniproxy's connection handling performance is comparable to that of full-fledged GNU/Linux TCP proxy implementations, but its minimalistic footprint enables new use cases. Specifically, Miniproxy requires as little as 6 MB to run and boots in tens of milliseconds, enabling massive consolidation, on-the-fly instantiation and edge cloud computing scenarios. We demonstrate the benefits of Miniproxy by implementing and evaluating a TCP acceleration use case.
\end{abstract}

\vspace{-0.1in}
\section{Introduction}
\label{sec:intro}
\input{intro}

\vspace{-0.1in}
\section{Related Work}
\label{sec:related}
\input{related}

\section{TCP acceleration}
\label{sec:tcp}
\input{reqs-arch}

\section{System overview}
\label{sec:system}
\input{system}

\section{Miniproxy Implementation}
\label{sec:implementation}
\input{implementation}

\section{Evaluation}
\label{sec:eval}
\input{evaluation}

\section{Conclusion}
\label{sec:conclusion}
\input{conclusion}

\small{
\bibliographystyle{abbrv}
\bibliography{bibliography-short2}
}

\end{document}

%% file: intro.tex
Service access times are directly correlated to users' experience and thus to service providers' revenues~\cite{singla2014internet}. Amazon estimates that an increase of delay of 100ms cuts its revenue by 1\%~\cite{Flach2013SIGCOMM}. Google measured a 0.74\% drop in the number of web searches performed by users when the search service delay was artificially increased by 400ms~\cite{Zhou2011CoNEXT}. Similarly, Bing experienced a reduction of 1.2\% in per-user revenue when the service delay was increased by 500ms~\cite{bingPerf}. Given the drastic impact that a few additional  milliseconds can have, network performance constitutes a critical element for many Internet services.

In today's networks, latency is dominated by two components: the round-trip time (RTT) between the communication's end-points and the number of RTTs required to complete the transfer~\cite{Flach2013SIGCOMM}. The RTT is determined by delay in the physical infrastructure, routing and queuing. In this work, we focus on reducing the number of RTTs required to complete a data transfer. In the case of TCP, optimizations have targeted most of the protocol mechanisms including connection establishment ~\cite{Radhakrishnan2011TFO, Ladiwala2009}, slow start~\cite{Al-Fares2011IMC} and congestion avoidance~\cite{Flach2013SIGCOMM}. However, since parts of TCP are fundamental to its correct operation, the optimization space is constrained or may require  extensive changes to the network infrastructure and protocol stack~\cite{Alizadeh2010SIGCOMM, Zhou2011CoNEXT}, making deployment harder.

A complementary approach to TCP optimization is the deployment of TCP proxies within the path of an end-to-end connection~\cite{liu2004}. A TCP proxy splits a single TCP connection into two connections and, if located in the middle (delay-wise) of the original connection's path, can noticeably speed up  end-to-end communication by reducing the feedback-loop time of each TCP connection~\cite{Ladiwala2009}. However, implementations of this optimization technique have been deployed only in cases in which the requirement for the optimization was known and well established in advance, as is the case for content distribution networks (CDNs)~\cite{akamai-statistics, Pathak2010PAM}. Dynamic, on-the-fly deployments of TCP proxies for connection acceleration have only been explored, so far, as a service provided by ``enhanced'' routers~\cite{Ladiwala2009}. Unfortunately, the deployment of such a solution is difficult, as it requires modification to the routers, and impractical, since it assumes that network flows do not undergo path changes due to routing throughout the lifetime of a flow.

Despite this state of affairs, recent trends are transforming the network into a cloud infrastructure that allows for running a variety of services in a number of different locations~\cite{MancoHotCloud15}; these provide new opportunities for on-the-fly deployment of proxies for TCP acceleration, at the right places in the network and in a timely fashion. To leverage these flexible cloud infrastructures, TCP proxies need to be virtualized, while still offering good performance and scalability.

In this paper we present \miniproxy, a lightweight, virtualized TCP proxy that can support scenarios such as on-the-fly TCP acceleration. \miniproxy is a Xen unikernel~\cite{unikernels}, is as fast as state-of-the-art GNU/Linux-based proxies, requires only 6 MB of RAM to run and can boot in just 12ms. Using it as a building block, we demonstrate that it is indeed possible to accelerate a TCP connection by deploying one or more proxies on an end-to-end path, even when placement of the proxies requires some deviation from the shortest path. Thanks to its small boot times, \miniproxy allows for the creation of chains of TCP proxies \emph{just-in-time}, by booting instances at locations in the network convenient for TCP acceleration.

 

%% file: related.tex
Our work follows the trend of using specialized VMs, also known as
unikernels~\cite{unikernels,clickos,erlangonxen,osv,mirage}, for
creating virtualized network functions. As opposed to previous work,
we focus on transparently accelerating TCP connections and
instantiating such servers on-the-fly. More recently,
Jitsu~\cite{jitsu} and the work in~\cite{in-net} proposed the
instantiation of VMs on-demand; we leverage similar mechanisms to show
that (virtual) TCP proxies can be instantiated just-in-time in order to
improve the performance of TCP flows.

Our work is also related to the vast literature regarding TCP acceleration. Specifically, we leverage the results of \cite{Ladiwala2009} and ~\cite{liu2004} to perform TCP acceleration, though we use a novel TCP proxy implementation, which can be dynamically deployed in virtualized network infrastructures. The work in~\cite{Radhakrishnan2011TFO} and \cite{Al-Fares2011IMC} optimizes the TCP connection establishment and slow start phases and so its results are complementary to ours. In \cite{Pathak2010PAM} TCP proxies are deployed in fixed locations to reduce the connection time; instead, we make the case for dynamically-placed proxies.

%% file: reqs-arch.tex
%
\setlength\textfloatsep{.5cm}


\begin{figure*}%
  \centering
  \begin{subfigure}{.65\columnwidth}
    \includegraphics[width=\columnwidth]{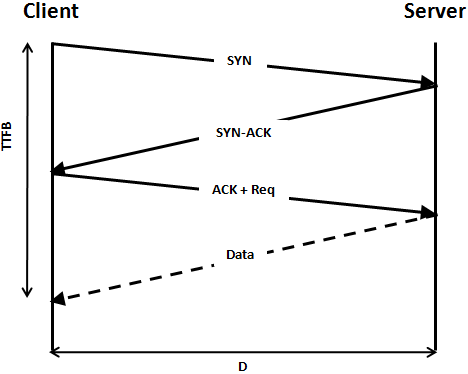}%
    \caption{3-way handshake}%
    \label{fig:ESF:a}%
  \end{subfigure}\hfill%
  \begin{subfigure}{.65\columnwidth}
    \includegraphics[width=\columnwidth]{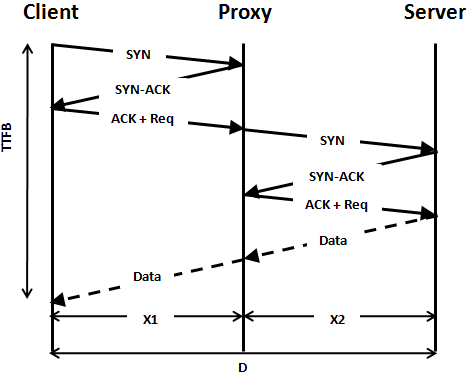}%
    \caption{3-way handshake and Proxy}%
    \label{fig:ESF:b}%
   \end{subfigure}\hfill%
  \begin{subfigure}{.65\columnwidth}
    \includegraphics[width=\columnwidth]{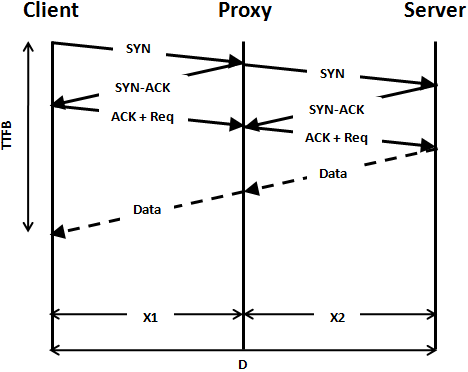}%
    \caption{Early SYN Forwarding}%
    \label{fig:ESF:c}%
   \end{subfigure}  %
   \caption{3-way handshake, 3-way handshake and Proxy, Early SYN Forwarding}
  \label{fig:ESF}
\end{figure*}

In general, a TCP connection comprises three phases: connection establishment (3-way handshake), slow start and congestion avoidance. After connection establishment, throughput increases exponentially during the slow start phase until it reaches a threshold, after which the congestion avoidance phase starts. For short flows, a TCP connection may terminate before reaching such threshold, which is the case for the large majority (about 90\%) of flows~\cite{Al-Fares2011IMC}; as a result, we mainly focus on optimizing the first two phases.

\vspace{0.1in}
\textbf{Connection Establishment}. 
TCP uses a 3-way handshake to establish a connection (see figure~\ref{fig:ESF:a}). For short flows, this procedure may constitute a significant part of the overall TCP flow's Time To Complete (TTC), i.e., the time it takes to open a connection, transfer all the data and close it. If the network delay between the client and the server is $D$ (assuming it is symmetric for simplicity), the 3-way handshake will take $3D$, assuming that the processing and transmission delay of the hosts are much smaller than the network delay: a SYN packet takes $D$ to get to the server, another $D$ is added for the server's SYN+ACK response to travel back to the client, and a final $D$ is required for the client's ACK to be finally received by the server. In addition, it is common to measure the Time To First Byte (TTFB) which includes the first bytes of data from the server to the client, that is, the TTFB is equal to 2 RTTs or $4D$.

For data transfers smaller than the TCP initial window (IW), the TTC is 2RTTs, that is, $TTC=TTFB$. For example, in the current implementation of GNU/Linux the IW is 10 segments (with each segment typically 1.5KB in size) and, assuming no packet loss, all the transfers smaller than 15KBs have $TTC=TTFB=2RTTs$.

Assume now a scenario with a TCP proxy. The client to server connection establishment requires now two different 3-way handshakes (see figure~\ref{fig:ESF:b}). Usually, the two 3-way handshakes are executed in sequence: the first one happens between the client and the proxy, and only after that the second one, between the proxy and the server, takes place. Assuming that $X_1$ is the client-proxy link delay, that $X_2$ is the proxy-server link delay, and that $X_1+X_2=D$, it can be shown that the TTFB does not change because of the proxy. In fact, the two 3-way handshakes take respectively $3X_1$ and $3X_2$, and $TTFB = 3X_1+3X_2+D = 4D$. The formula can be generalized for a chain of $N$ proxies ($N+1$ 3-way handshakes): assuming an even split of the end-to-end delay ($X = D/(N+1)$), each 3-way handshake takes $3*X = 3*D/(N+1)$, and the TTFB is $(N+1)*3*D/(N+1)+D = 4D$.

Adopting the technique proposed in \cite{Ladiwala2009}, TCP proxies can speedup the connection establishment. In this approach, the client's TCP SYN packet is forwarded to the next TCP hop as soon as it is received by a proxy (see figure~\ref{fig:ESF:c}). We refer to this mechanism as \emph{Early SYN Forwarding} (ESF). With ESF, the two 3-way handshakes are partially executed in parallel, reducing the overall connection establishment time and TTFB. Consider again the single proxy scenario: the SYN arrives to the server with a delay $(X_1+X_2)=D$, but the SYN-ACK arrives to the client at $2*X_1$, so that the client's ACK is received by the proxy already at $3*X_1$. Thus, the ACK can be sent by the proxy to the TCP server at $max(3*X_1;X_1+2*X_2)=X_1+2*max(X_1;X_2)$ and arrives to the TCP server with an additional delay $X_2$. Therefore, the 3-way handshake takes
\vspace{-0.02in}
\begin{flalign} 
\nonumber X_1+X_2+2*max(X_1;X_2)=D+2*max(X_1;X_2)
\end{flalign}

\noindent and the TTFB takes an additional $D$:
\vspace{-0.02in}
\begin{flalign} \label{eq:ttfb_2hops}
TTFB &=2D+2*max(X_1;X_2) &&
\end{flalign}

If $X_1=X_2$, we are in the best case with $TTFB=3D$, saving the 25\% of the original TTFB between client and server. Generalizing with N proxies, under the optimal assumption that the delay $X$ on each of the $N+1$ links is the same $X = D/(N+1)$, then the 3-way handshake is completed in $D+2X=D+2D/(N+1)$, while the TTFB is $2D+2D/(N+1)$. In case of different delays $X_i$ for the N+1 hops, eq.(\ref{eq:ttfb_2hops}) can be generalized for an arbitrary number of hops as follows: 
\begin{flalign} \label{eq:ttfb_any_hops}
TTFB = 2D+2*max(X_i) &&
\end{flalign}



\textbf{Slow start}
The slow start phase can be modeled by considering a sequence of time slots of duration RTT=2D [s], numbered with $k$. In the first slot ($k=0$) a number of segments equal to $IW$ (Initial Window) is sent. Considering the exponential increase, in a generic slot $k$ during the slow start phase the number of sent segments is $2^k*IW$. Let $S_k$ and $B_k$ be the number of segments and bytes sent in the slot $k$, $R_k$ the throughput during slot $k$, $Btot_k$ the cumulative amount of data that can be transferred up to the slot $k$ (included), $Ra_k$ the average throughput up to slot $k$ (included), $TTC_k$ the time to complete for an amount of data $Btot_k$. We can evaluate these metrics as follows:

\setlength{\belowdisplayskip}{0pt} \setlength{\belowdisplayshortskip}{0pt}
\setlength{\abovedisplayskip}{0pt} \setlength{\abovedisplayshortskip}{0pt}

\begin{flalign} \label{eq:ss1}
S_k &= 2^k*IW &&
\end{flalign}
\begin{flalign} \label{eq:ss2}
B_k &=2^k*IW*MSS \; [bytes] &&
\end{flalign}
\begin{flalign} \label{eq:ss3}
R_k &=\frac{B_k*8}{RTT} = \frac{2^k*IW*MSS*8}{2D}  \; [b/s] &&
\end{flalign}
\begin{flalign} \label{eq:ss4}
Btot_k &=\sum\limits_{i=0}^k B_i \; [bytes] &&
\end{flalign}

\begin{flalign} \label{eq:ss5}
Ra_k= \frac{1}{k+1} \sum\limits_{i=0}^k R_{i} =\frac{IW*MSS*8}{RTT}\frac{(2^{k+1}-1)}{k+1} \;[b/s] &&
\end{flalign}
\begin{flalign} \label{eq:ttc_k} 
TTC_k = TTFB + k*RTT = TTFB + k*2D \; [s] &&
\end{flalign}
\\
If the regular 3-way handshake is used (i.e., no ESF) $TTFB = 4D$, we have:\\
\begin{flalign} \label{eq:ttc_k_noESF}
TTC_k = 2D + 2D + k*2D \;[s] &&
\end{flalign}
\\
Eq.(\ref{eq:ttc_k_noESF}) decomposes $TTC_k$ in three parts: the first one represents the RTT that cannot be reduced, the second one can be reduced with TCP proxies and the ESF mechanism as shown in eq.(\ref{eq:ttfb_any_hops}), and the third one is reduced with TCP proxies as shown hereafter in eq.(\ref{eq:ttc_k_any_hops}). 

%

According to eq.(\ref{eq:ttc_k}), for a given amount of data, the duration of the slow start phase is directly proportional to RTT. Introducing a proxy in the path splits the TCP connection in two. In each of the split connections, the slow start phase starts when the first data segment is received. The data segment is immediately forwarded downstream so that the slow start phases in each part proceed in parallel. If the proxy splits $D$ in even parts, the duration of the slow start phase is halved. Considering $N$ proxies and $N+1$ hops with arbitrary delays $X_i$, the general expression of $TTC_k$ is: 

\begin{flalign} \label{eq:ttc_k_any_hops}
\nonumber TTC_k &= TTFB+k*2*max(X_i) \; [s] && \\
&= 2D + 2*max(X_i) + k* 2*max(X_i) && \\\nonumber 
\end{flalign}


If all the delays $X_i = D/(N+1)$, eq.(\ref{eq:ttc_k_any_hops}) becomes:\\ 
\begin{flalign} \label{eq:ttc_k_equal_hops}
TTC_k = 2D + 2D/(N+1) + k* 2D/(N+1) \; [s] &&
\end{flalign}
\\



%% file: system.tex
In this section, we present an overview of a system that leverages Miniproxy to provide on-the-fly TCP acceleration. Although this paper is mainly focused on the implementation of the Miniproxy itself, we believe a description of one of its applications will provide a clearer understanding of the advantages of the proposed technology.

The aim of the envisioned system is to accelerate an end-to-end TCP connection by deploying a chain of TCP proxies. 
For such an objective, it is critical to properly locate the proxies. In fact, split TCP achieves best performance when minimizing the overall end-to-end RTT (i.e., the sum of the individual connections' RTTs) with an even distribution of the connections' RTTs. 
However, the best locations are strictly dependent on the actual end-points relative locations. That is, each end-to-end connection has its own best proxies' locations. Furthermore, for any given couple of end-points, those locations may need to be changed over time because of the changing traffic conditions. In the light of these considerations, it seems evident that a network provider is the best suited actor to provide such a service. However, given that the TCP flows being more suitable for acceleration are those that experience large RTTs, it is neat to assume that the proxies' locations may be distributed among several network providers. Hence it seems difficult to deploy this acceleration technique as a network infrastructure service~\cite{Ladiwala2009}.

Luckily, today's Internet offers a set of locations for flexibly deploying TCP proxies: publicly available cloud datacenters. Providers such as Amazon, Microsoft and Google are just the most well-known ones, but a number of national and regional providers can be easily added to the list. 
Using the cloud to run proxies does not just solve the locations issue, but actually includes a new variable in the picture. Cloud providers offer a flexible utility-based approach to run VMs, with a fine-grained (sometimes per-minute of activity) billing model. Clearly, such a cost model strongly calls for a system that is able to run a VM only for the required amount of time, when it is actually required. 

\vspace{0.1in}
\noindent \textbf{Requirements}. 
Achieving such objectives hinges on fulfilling a number of requirements. 
First, we would like to leverage the availability of cloud deployments and virtualized infrastructure in order to ease deployment and obtain the best acceleration performance out of our proxies; consequently, the proxies should be virtualized in order to run over such platforms.
Second, using cloud deployments requires minimizing the time each proxy (i.e., VM) runs, in order to lower the costs of running such instances when employing a utility cost model. For example, this suggests that proactively deploying the proxies may not be an optimal solution from a cost perspective. Furthermore, for new TCP connections, at least one of the TCP connection's end-points is unknown before the connection is initiated; thus, it is not possible to know, in advance, the best locations for the proxies. Since our gains can only come from the connection establishment and slow start phases, a third crucial requirement is to be able to place the proxies at their optimal locations in very short timescales. 
Third, each pair of end-points may potentially require different proxy locations and those locations can change over time due to varying network conditions; this may require running many proxies at many locations, flexibly scaling their numbers, to optimally accelerate a set of connections. 
Finally, a big number of proxies means that proxy instances should be as lightweight as possible to reduce the cost of running them.

\vspace{0.1in}
\noindent \textbf{Building blocks}.
The implementation of a system that fulfills the above requirements is technically challenging and requires a number of building blocks. For instance, a building block is a monitoring system that checks the network conditions between the suitable locations; the collected information would be then required by another building block, e.g., an orchestration system, which uses this information to select the best locations to run the proxies. It is also required to design and implement the system that manages and timely creates the proxies, as well as the mechanisms to forward packets to the proxy VMs in the hypervisors and to chain together several proxies. 

This list could be actually much longer, and it is out of the scope of this work to describe a full-fledged solution. Instead, in the next section we focus on the basic building block for such a system, i.e., a virtualized TCP proxy.

%% file: implementation.tex
As shown in the previous section, the best performance is guaranteed by minimizing the overall end-to-end RTT (i.e., the sum of the individual connections' RTTs) with an even distribution of the connections' RTTs. Achieving such an objective hinges on fulfilling a number of requirements. 

The first one arises from the fact that at least one of the TCP connection's end-points is unknown before the connection is initiated; thus, it is not possible to know, in advance, the  locations for the proxies. Since our gains can only come from the connection establishment and slow start phases, it is crucial to be able to place the proxies at their optimal locations in very short timescales.

Second, each pair of end-points may potentially require different proxy locations and those locations can change over time due to varying network conditions; this may require running many proxies at many locations to optimally accelerate a set of connections. Third, a big number of proxies means that proxy instances should be as lightweight as possible to reduce the cost of running them. Finally, as discussed in the introduction, we would like to leverage the availability of cloud deployments and virtualized infrastructure in order to ease deployment and obtain the best acceleration performance out of our proxies; consequently, the proxies should be virtualized in order to run over such platforms.

\vspace{0.1in}
\textbf{Implementation}.
Given the above requirements, we would like to leverage the nice
properties of VMs such as isolation but without incurring 
their overheads. To do so, we settle on unikernels~\cite{unikernels}:
purpose-built, specialized VMs based on minimalistic
OSes. Unikernels have a number of advantages including a single
address space, so no expensive system calls; low memory footprint (MBs
or even KBs) and fast instantiation times (milliseconds compared to
seconds for conventional VMs). These properties are important since
they let us instantiate our TCP proxies on demand, and potentially
a large number of them on the same box.

\begin{figure}
  \begin{center}
     \includegraphics[width=1\linewidth]{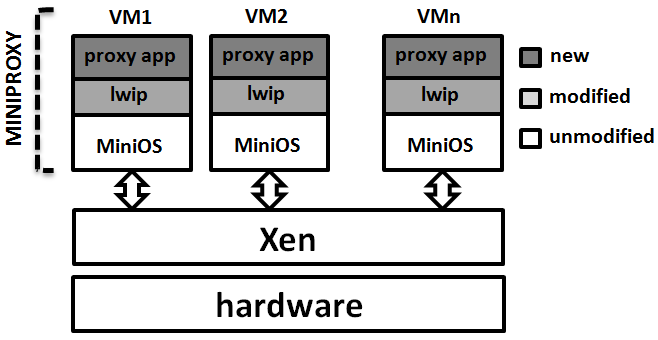}

    \vspace{0.1in}
    \caption{\miniproxy architecture showing modified and new components.}
    \label{fig:arch}
  \end{center}
\end{figure}

In particular, we target unikernels on Xen, and use the paravirtualized MiniOS~\cite{minios} operating system to build on (see figure~\ref{fig:arch}). To implement \miniproxy, we make a number of modifications to \texttt{lwip} (a small, open source TCP/IP stack for embedded systems) to handle the actual TCP connections, and develop a MiniOS-based proxy application from scratch that leverages these modifications. We further make use of the optimizations to Xen and MiniOS described in~\cite{MancoHotCloud15} in order to derive even smaller boot times.

In greater detail, we modify \texttt{lwip} to model a TCP proxy as pairs of sockets, one for the connection between the client (or previous hop) and the proxy and one for the connection between the proxy and the server (or next hop). In other words, the first socket is used to receive incoming connections while the second one is used to open a connection towards the server (or the next hop). To implement this in \texttt{lwip}, we linked together two PCBs (Protocol Control Block) structures. Our API allows us to instantiate a PCB pair through a new bind function called \texttt{tcp\_{}early\_{}syn\_{}bind()}. When a PCB pair is instantiated, each PCB plays a specific role: one listens for incoming connections, while the other one is ready to forward the SYN packet (\texttt{outgoing\_{}idle} state). Each of the PCBs performs a separate three-way connection establishment process and invokes an application callback when done. The listening PCB hands over information about the connection with the endpoint that has started the communication; likewise, the \texttt{outgoing\_{}idle} PCB includes state to support communication with the target endpoint. 

In addition to this change, we added a new callback to \texttt{lwip} that is invoked when the SYN is received, allowing applications to instantiate data structures and take decisions about the next hop, if needed. We further increase the TCP IW to 10 segments, the same value used by the majority of the TCP implementations. We set the TCP send buffer size, configurable by the user, to double the TCP receive window, as this resulted in the best performance in our experimental tests. 

The other major change to \emph{lwip} has to do with the introduction of a new TCP option that allows us to chain together a set of explicit proxies. Briefly, there are two ways to send traffic through proxies: implicit (or transparent)~\cite{Ladiwala2009} and explicit~\cite{liu2004}. In the implicit case the flow's destination can be inferred from packet headers, i.e., destination IP address and TCP port, which means that the proxy must be on the path between the client and the packet's IP destination. The advantage is that all the information to identify the final end-point of the TCP connection is in the packet's header. In the explicit case, instead, the IP destination of the TCP connection is the proxy itself, meaning that the proxy has to read the TCP flow's data to learn the final destination of the connection. For example, when using HTTP proxies, the HTTP header includes the destination server in the URI. Unfortunately, while explicit proxies can be deployed anywhere, they do not allow for ESF (Early SYN forwarding), since they need to complete the first 3-way handshake to read the TCP flow's data. To overcome this limitation, we introduce a new TCP option that carries information about the source/destination IP addresses and TCP ports inside a SYN packet. The use of this option extends the support for ESF also to explicit proxies. 

The TCP option can be introduced directly by the client or by an implicit proxy which is on-path and can redirect the connection to an explicit proxy. Note that at least one implicit proxy on the path is already commonly deployed, e.g., in cellular networks~\cite{Wang2011SIGCOMM, XuPAM15a}. In this case, the first proxy (likely close to the client), will perform a traditional 3-way handshake, while all the remaining proxies in the chain, if supporting the newly introduced TCP option, can use ESF. \miniproxy supports the new TCP Option, which contains the 4-tuple IP source/destination address and TCP source/destination port. 


The implementation of the actual proxy application running on top of MiniOS and our modified \texttt{lwip} stack consists of 600 LoC, is event-driven, requires a minimum of 6MB of RAM, and does not need any block devices. Besides performing the TCP acceleration already described, the proxy is also able to parse the new TCP option and  apply per-flow policies. Finally, it is worth pointing out that in the current state \miniproxy performs only TCP proxying, but it has been designed for extensibility in order to eventually introduce additional functions, e.g., caching.

%% file: evaluation.tex
%
\setlength\textfloatsep{.5cm}

In this section we present a preliminary performance evaluation of our \miniproxy implementation.
We first perform micro-benchmarks of \miniproxy's (1) throughput, (2) connection establishment time and (3) boot times, followed by an evaluation of how applying ESF and a chain of proxies affects TCP performance. Each measurement presented in this section is the average across several runs (100). Confidence intervals are not plotted since they are very close to the average.

\subsection{Micro-benchmarks}

Unless otherwise stated, all tests in this section were run on a server with an Intel Xeon CPU @3.4GHz, 16GB of RAM and a dual port Intel x540 10Gb NIC on Xen 4.4. The server is connected back-to-back to a traffic generator server. Traffic is generated there, forwarded by the box running \miniproxy and sent back to the generator server, which then measures throughput.

\noindent\textbf{Throughput}: For the first experiment, we measured the throughput achieved by \miniproxy~ and compared it to the one achieved by Varnish, a state-of-the-art TCP proxy implementation for GNU/Linux. In this test \miniproxy~ is configured to run with 8 MB of RAM, while Varnish requires 1 GB. Both proxies run on a VM to which we dedicate a single CPU core, and we instrument the generator to send a single TCP flow. With this setup, \miniproxy consistently outperforms Varnish in terms of throughput by about 5\% (1.534 Gb/s versus 1.462 Gb/s for Varnish) while consuming significantly less memory. It is worth noting that the results were without  Generic Segmentation Offload (GSO~\cite{gso}): we have already added GSO support to \texttt{lwip} and early numbers, without using the proxy code, are encouraging (in the range of 32 Gb/s). We are now in the process of modifying our proxy application to take advantage of GSO.

\begin{table}[ht]
\centering 
\begin{tabular}{c c c c c c c} 
\hline\hline 
\# Concurrent Conn.  & 30 & 70 & 110 & 150 & 190 & 230  \\
\hline 
Avg. time (ms) & 0,1 & 1 & 2 & 2 & 3 & 3  \\ [1ex] 
\hline 
\end{tabular}
\caption{Connection establishment time} 
\label{table:conn} 
\end{table}

\noindent\textbf{Connection Establishment}: We measure \miniproxy's connection establishment time when varying the number of simultaneous connections (see table~\ref{table:conn}). Our test shows that \miniproxy is able to handle an increasing number of simultaneous connections without overly increasing the per-flow connection establishment time (in the worst case we measure a 3ms establishment time when handling 230 simultaneous connection requests). 

\begin{figure}[htb!]
  \begin{center}
    \includegraphics[width=1\linewidth]{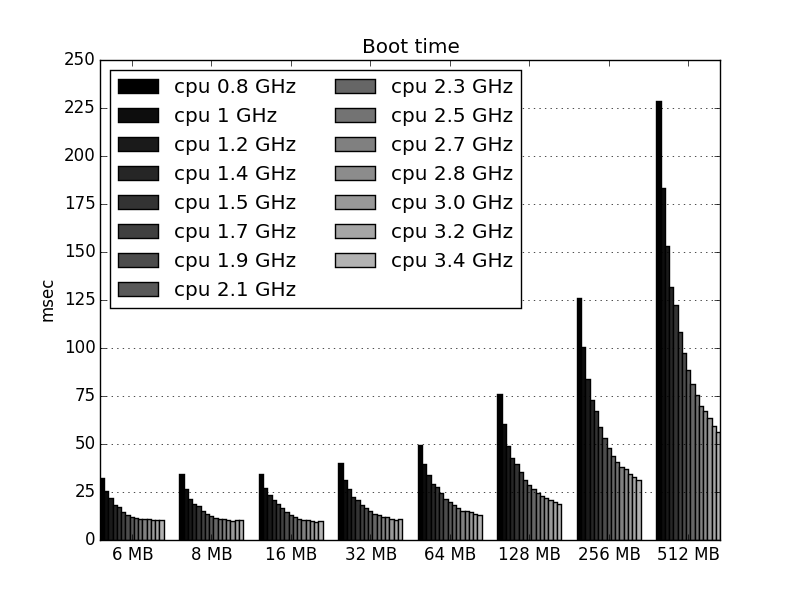}
    \caption{Miniproxy boot times for different CPU frequencies.}
    \label{fig:boottime}
  \end{center}
\end{figure}

\noindent\textbf{Instantiation Time}: Finally, we measured the boot time of a \miniproxy instance when varying both the amount of RAM allocated to the VM and the CPU frequency. In this test, the RAM size directly impacts the number of simultaneous flows the proxy can handle. The evaluation for different CPU frequencies, instead, is useful in order to show that \miniproxy can run on devices with different capabilities, e.g., \miniproxy could be deployed on home gateways at the edge of the network or on big servers in cloud datacenters. Figure ~\ref{fig:boottime} shows the results. \miniproxy requires a minimum of 6 MB to boot, in which case the instance boots in just 12ms for CPU frequencies above 2GHz. For a CPU frequency of 0.8GHz (the minimum our system allows), \miniproxy~ is still able to boot in just 30ms. For bigger RAM amounts, the boot times go up to 60ms on a 3GHz CPU and 230ms on a 0.8GHz CPU. 

\subsection{TCP Acceleration with ESF}
Having shown that \miniproxy can meet the requirements mentioned in section~\ref{sec:implementation}, we now perform a number of tests to verify to which extent we can use it to accelerate TCP connections. In these experiments we assume a RTT of 100ms between client and server~\cite{Al-Fares2011IMC}, a link bandwidth of 100Mb/s and that the proxies apply ESF and evenly split the end-to-end RTT.
The delays are generated synthetically using the Linux Network Emulation (netem) tool.


\begin{figure}[htb!]
  \begin{center}
    \includegraphics[width=1\linewidth]{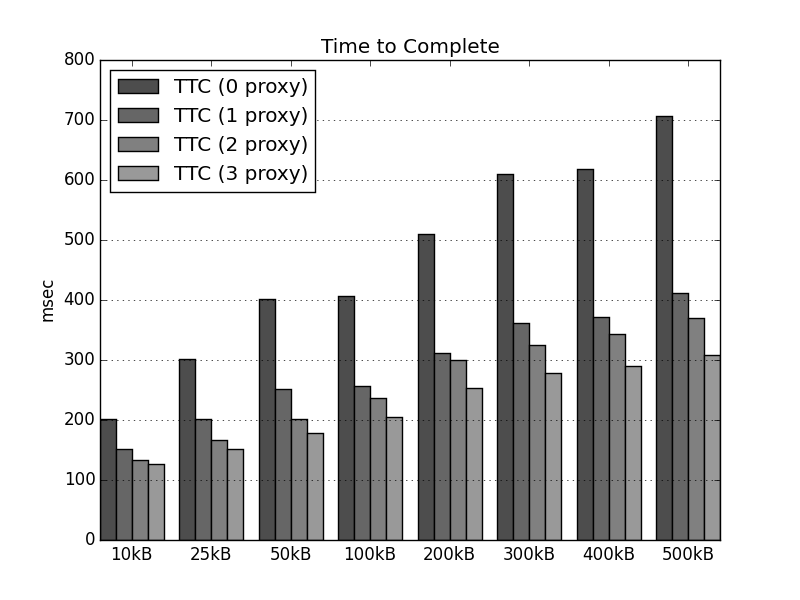}
      \caption{Transfer time when varying the number of proxies for different transfer sizes.}
    \label{fig:tt}
  \end{center}
\end{figure}

Figure~\ref{fig:tt} shows the Time To Complete (TTC) for different flow sizes (from 10 KBs to 500 KBs) using a variable number of proxies. For TCP flows of 10 KBs we have $TTC=TTFB$, so we are in fact measuring the performance of the ESF implementation. The ESF mechanism yields a TTFB reduction of 25\%, 33.3\% and 37.5\% when using 1, 2 and 3 proxies, respectively. This result matches the theoretical model discussed in section~\ref{sec:tcp}. 

For flows that need more than 2 RTTs to complete, we can also see the effects of a faster slow start phase due to the shorter RTT. With respect to the TTC achieved without a proxy, the TTC for 25KB flows is improved by an additional 8\%, 10\% and 11,5\% due to the increased throughput in the slow start phase. In total, we measured a reduction of 33\%, 44\% and 49\% in TTC when using 1, 2 and 3 proxies, respectively. Of course, the longer the duration of the slow start phase, the better the measured acceleration. 

\begin{figure}
  \begin{center}
    \includegraphics[width=1\linewidth,clip]{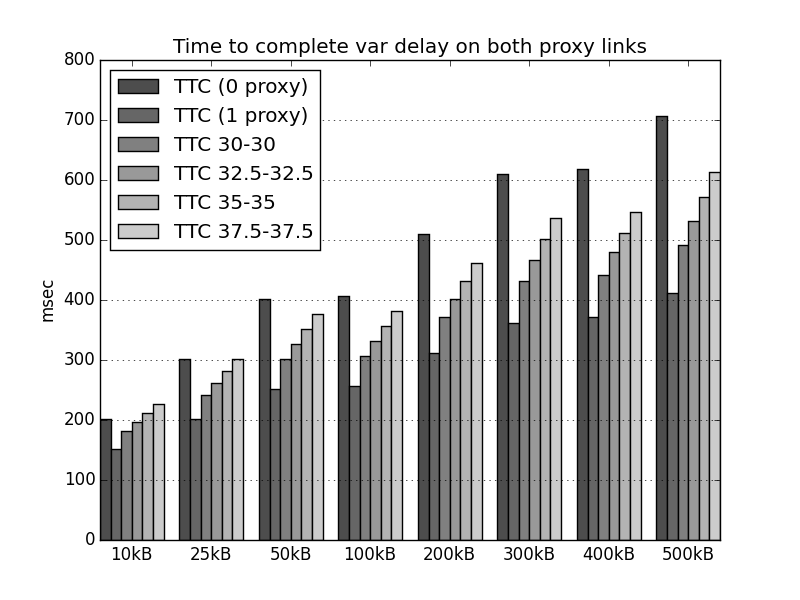}
    \caption{Transfer time with one proxy for different transfer sizes. The proxy introduces additional delay on the client-proxy and proxy-server links.}
    \label{fig:tt-cdpds}
  \end{center}

\end{figure}

An interesting case is when a proxy is located off-path, which introduces additional delay. To quantify this, we perform a test with an RTT of 100 ms (the one way delay is 50ms) and start from the ideal case of an on-path proxy that splits the one way delay in two segments of 25ms each. We then increase the delay on both links from 25 to 37.5ms (in steps of 2.5ms). The introduction of a proxy that splits the connection but increases the end-to-end delay D can produce different results, as explained by eqs.(\ref{eq:ttc_k_noESF}) and (\ref{eq:ttc_k_any_hops}). For small increases in D, there is an advantage in the split because $D$ is replaced by $max(X_i)$ in the second and third terms of eq.(\ref{eq:ttc_k_noESF}). When the increase of $D$ is large (and consequently also $max(X_i)$ increases) the TTC increases. Figure~\ref{fig:tt-cdpds} illustrates the effect of the longest delay $max(X_i)$, and shows that in this specific case, even an increase of 50ms in the overall RTT still produces a reduction of the TTC when compared to the case with no proxies (for flows that require at least 2 RTTs to complete).

%% file: conclusion.tex
We presented \miniproxy, a lightweight TCP proxy that can boot in 12ms, while providing high forwarding performance. We showed that \miniproxy~can be used to accelerate TCP's connection establishment and slow start phases. \miniproxy's very short boot times enable us to even boot the proxy instances on-the-fly, as the SYN packet of a TCP connection is first received.

\miniproxy is so far a simple proof-of-concept. Going forward, we intend to develop strategies for optimally selecting where to boot \miniproxy instances given a flow's endpoints and a set of available deployment locations. Further, we are also extending our evaluation to include wide area experiments, taking into account different traffic conditions and evaluating the suitability of available cloud datacenters (e.g., Amazon's EC2) as \miniproxy execution environments.

%% file: main.bbl
\begin{thebibliography}{10}

\bibitem{akamai-statistics}
Akamai.
\newblock Akamai's state of the internet: Q1 2015 report.

\bibitem{Al-Fares2011IMC}
M.~Al-Fares et~al.
\newblock Overclocking the yahoo!: Cdn for faster web page loads.
\newblock In {\em Proceedings of the 2011 ACM SIGCOMM Conference on Internet
  Measurement Conference}, IMC '11, pages 569--584, New York, NY, USA, 2011.
  ACM.

\bibitem{Alizadeh2010SIGCOMM}
M.~Alizadeh et~al.
\newblock Data center tcp (dctcp).
\newblock In {\em Proceedings of the ACM SIGCOMM 2010 Conference}, SIGCOMM '10,
  pages 63--74, New York, NY, USA, 2010. ACM.

\bibitem{erlangonxen}
{Erlang on Xen}.
\newblock {Erlang on Xen}.
\newblock \url{http://erlangonxen.org/}, {July} {2012}.

\bibitem{Flach2013SIGCOMM}
T.~Flach et~al.
\newblock Reducing web latency: The virtue of gentle aggression.
\newblock In {\em Proceedings of the ACM SIGCOMM 2013 Conference on SIGCOMM},
  SIGCOMM '13, pages 159--170, New York, NY, USA, 2013. ACM.

\bibitem{osv}
A.~Kivity et~al.
\newblock Osv{\textemdash}optimizing the operating system for virtual machines.
\newblock In {\em 2014 USENIX Annual Technical Conference (USENIX ATC 14)},
  pages 61--72, Philadelphia, PA, June 2014. USENIX Association.

\bibitem{Ladiwala2009}
S.~Ladiwala et~al.
\newblock Transparent tcp acceleration.
\newblock {\em Comput. Commun.}, 32(4):691--702, Mar. 2009.

\bibitem{liu2004}
Y.~Liu, Y.~Gu, H.~Zhang, W.~Gong, and D.~Towsley.
\newblock Application level relay for high-bandwidth data transport.
\newblock {\em Proc. of GridNets}, 2004.

\bibitem{mirage}
A.~Madhavapeddy et~al.
\newblock Unikernels: library operating systems for the cloud.
\newblock In {\em Proceedings of the eighteenth international conference on
  Architectural support for programming languages and operating systems},
  ASPLOS '13, pages 461--472, New York, NY, USA, 2013. ACM.

\bibitem{jitsu}
A.~Madhavapeddy, T.~Leonard, M.~Skjegstad, T.~Gazagnaire, D.~Sheets, D.~Scott,
  R.~Mortier, A.~Chaudhry, B.~Singh, J.~Ludlam, J.~Crowcroft, and I.~Leslie.
\newblock {Jitsu: Just-In-Time Summoning of Unikernels}.
\newblock In {\em NSDI}, 2015.

\bibitem{MancoHotCloud15}
F.~Manco, J.~Martins, K.~Yasukata, J.~Mendes, S.~Kuenzer, and F.~Huici.
\newblock The case for the superfluid cloud.
\newblock In {\em Proceedings of the 7th USENIX Conference on Hot Topics in
  Cloud Computing}, HotCloud'15, pages 7--7, Berkeley, CA, USA, 2015. USENIX
  Association.

\bibitem{clickos}
J.~Martins, M.~Ahmed, C.~Raiciu, and F.~e. Huici.
\newblock Enabling fast, dynamic network processing with clickos.
\newblock In {\em Proceedings of the second ACM SIGCOMM workshop on Hot topics
  in s\ oftware defined networking}, HotSDN '13, pages 67--72, New York, NY,
  USA, 2013. ACM.

\bibitem{Pathak2010PAM}
A.~Pathak et~al.
\newblock Measuring and evaluating tcp splitting for cloud services.
\newblock In {\em Proceedings of the 11th International Conference on Passive
  and Active Measurement}, PAM'10, pages 41--50, Berlin, Heidelberg, 2010.
  Springer-Verlag.

\bibitem{Radhakrishnan2011TFO}
S.~Radhakrishnan et~al.
\newblock Tcp fast open.
\newblock In {\em Proceedings of the Seventh COnference on Emerging Networking
  EXperiments and Technologies}, CoNEXT '11, pages 21:1--21:12, New York, NY,
  USA, 2011. ACM.

\bibitem{singla2014internet}
A.~Singla et~al.
\newblock The internet at the speed of light.
\newblock In {\em Proceedings of the 13th ACM Workshop on Hot Topics in
  Networks}, page~1. ACM, 2014.

\bibitem{bingPerf}
S.~Souders.
\newblock Velocity and the bottom line.
\newblock
  \url{http://radar.oreilly.com/2009/07/velocity-making-your-site-fast.html }.

\bibitem{in-net}
R.~Stoenescu, V.~Olteanu, M.~Popovici, M.~Ahmed, J.~Martins, R.~Bifulco,
  F.~Manco, F.~Huici, G.~Smaragdakis, M.~Handley, and C.~Raiciu.
\newblock In-net: In-network processing for the masses.
\newblock In {\em Proceedings of the European conference on Computer systems},
  EuroSys '15. ACM, 2015.

\bibitem{Wang2011SIGCOMM}
Z.~Wang et~al.
\newblock An untold story of middleboxes in cellular networks.
\newblock In {\em Proceedings of the ACM SIGCOMM 2011 Conference}, SIGCOMM '11,
  pages 374--385, New York, NY, USA, 2011. ACM.

\bibitem{unikernels}
{Xen Project}.
\newblock {The Next Generation Cloud: The Rise of the Unikernel}.
\newblock
  \url{http://wiki.xenproject.org/mediawiki/images/3/34/XenProject_Unikernel_Whitepaper_2015_FINAL.pdf},
  April 2015.

\bibitem{minios}
{Xen.org}.
\newblock {Mini-OS}.
\newblock \url{http://wiki.xen.org/wiki/Mini-OS}, {2015}.

\bibitem{gso}
H.~Xu.
\newblock Gso: Generic segmentation offload.
\newblock \url{https://lwn.net/Articles/188489/}, {2006}.

\bibitem{XuPAM15a}
X.~Xu et~al.
\newblock {Investigating Transparent Web Proxies in Cellular Networks}.
\newblock In {\em PAM}. Springer, 2015.

\bibitem{Zhou2011CoNEXT}
W.~Zhou et~al.
\newblock Asap: A low-latency transport layer.
\newblock In {\em Proceedings of the Seventh COnference on Emerging Networking
  EXperiments and Technologies}, CoNEXT '11, pages 20:1--20:12, New York, NY,
  USA, 2011. ACM.

\end{thebibliography}
